\documentclass[twocolumn,showpacs,showkeys,preprintnumbers,amsmath,amssymb]{revtex4}

\usepackage{psfig}
\usepackage{epsf}
\usepackage{graphicx}
\usepackage{dcolumn}
\usepackage{bm}

\begin{document}


\title{CO adsorption on Cu(211) surface: first principle and STM study.}

\author{Marek Gajdo\v s}\email{marek.gajdos@univie.ac.at}

\author{Andreas Eichler}%
\author{J\"urgen Hafner}%
\affiliation{%
Institut f\"ur Materialphysik and
Center for Computational Materials Science,\\
Universit\"at Wien, Sensengasse 8/12, A-1090 Wien, \"Osterreich
}%

\author{Gerhard Meyer}%
\affiliation{%
IBM Research, Zurich Research Laboratory, CH-8803 R\"uschlikon, Switzerland
}%

\author{Karl-Heinz Rieder}%
\email{rieder@physik.fu-berlin.de}
\affiliation{%
Institut f\"ur Experimentalphysik, Freie Universit\"at Berlin, Arnimallee 14, D-14195 Deutschland
}%

\date{\today}

\begin{abstract}
Chemisorption of CO on the stepped Cu(211) surface is studied within ab-initio density functional theory (DFT) and scanning tunneling microscopy (STM) imaging as well as manipulation experiments. Theoretically we focus on the experimentally observed ordered (2$\times$1) and (3$\times$1) CO-phases at coverages $\Theta$~=~$\frac{1}{3}$, $\frac{1}{2}$ and $\frac{2}{3}$~monolayer (ML). To obtain also information for isolated CO molecules found randomly distributed at low coverages, we also performed calculations for a hypothetical (3$\times$1) phase with $\Theta$~=~$\frac{1}{3}$ ML.  The adsorption geometry, the stretching frequencies, the work functions and adsorption energies of the CO molecules in the different phases are presented and compared to experimental data. Initially and up to a coverage of $\frac{1}{2}$ ML CO adsorbs upright on the on-top sites at step edge atoms. Determining the most favorable adsorption geometry for the $\frac{2}{3}$ ML ordered phase turned out to be nontrivial, both from the experimental and the theoretical point of view. Experimentally, both top-bridge and top-top configurations were reported, whereby only the top-top arrangement was firmly established. The calculated adsorption energies and the stretching frequencies favor the top-bridge configuration. The possible existence of both configurations at $\frac{2}{3}$ ML is critically discussed on the basis of the presently accessible experimental and theoretical data.
In addition, we present observations of STM manipulation experiments and corresponding theoretical results, which show that CO adsorbed on-top of  a single Cu-adatom, which is manipulated to a location close to the lower step edge, is more strongly bound than CO on-top of a step edge atom.
\end{abstract}

\maketitle

\vspace{0.2cm}
\section{Introduction}
\label{Introduction}
Experimental and theoretical studies of chemisorption on vicinal surfaces are of great current interest due to close connection to real catalytic substrates which exhibit beside other defects many steps. Thanks to scanning tunneling microscopy (STM) a breakthrough in determining the morphology of substrates and overlayers was achieved in recent years \cite{rmp:Binning:71}. Carbon monoxide on Cu(211) was studied with respect to ordered overlayers and also in connection with the possibility to laterally manipulate adsorbates on the surface by STM \cite{prl:Meyer:77,cpl:Bartels:273}. Meyer et al. \cite{cpl:Meyer:240} found that the Cu(211) surface does not reconstruct and that there are several ordered CO structures upon adsorption. We concentrate here on the (2$\times$1) and the (3$\times$1) phases with respective coverages of $\Theta$=$\frac{1}{3}$, $\frac{1}{2}$ and $\frac{2}{3}$~ML. STM imaging of disordered adsorption at very low coverage and of the (2$\times$1) phase revealed that the CO molecules adsorb on top of Cu atoms at the step edges. The high coverage (3$\times$1) phase is more complicated: upon preparing an adlayer at $100$~K, the admolecule configuration appears to be top-bridge, whereas in small islands prepared at $15$~K by lateral manipulation of the molecules they occupy top-top arrangements within the (3$\times$1) unit cell \cite{cpl:Zophel:310}. 

Braun et al. \cite{jcp:Braun:105} have observed a significant decrease in the energy of the frustrated translation vibrations parallel to the surface (T-mode) of isolated CO molecules adsorbed on the intrinsic step sites of Cu(211) as compared to CO on flat Cu(111) and Cu(001) surfaces. They propose a tilting of the CO molecules due to the presence of these intrinsic steps as a possible explanation for this energy decrease.

There have been several theoretical investigations of the geometric structure of clean Cu(211) \cite{prb:Wei:57,prb:Geng:64} and also Cu(n11) surfaces \cite{prb:Tian:47,ss:Spisak:489}. Very recently, reviews of electronic, structural, vibrational and thermodynamic characterization of vicinal surfaces were published \cite{jpcm:Mugarza:15,jpcm:Rahman:15,jpcm:Barreteau:15}. On the other hand, calculations on the adsorption of CO molecules on Cu(211) are less numerous \cite{ss:Rouzo:415}. Furthermore, theoretical investigations of vibrational and energetic properties  of the CO adsorption system are scarce or missing. Adsorption of CO on Cu(211) is also interesting from the fundamental point of view as it offers an occasion to study adsorbate reorganization as proposed for NO on the Pd(211) surface \cite{prl:Hammer:79}. 

In this work we carried out a systematic theoretical investigation of the clean copper (211) surface and of the adsorption of CO molecules. In section \ref{Clean Cu(211) surface} we examine the clean Cu(211) surface (multilayer relaxations, work-functions). The remainder of the paper is organized as follows. Section \ref{CO adsorption} is divided into several parts where we describe structural, vibrational and energetic properties of the CO molecules adsorbed on the Cu(211) surface. Moreover, subsection \ref{STM pictures} provides calculated STM images of the probable structural phases at $\Theta=\frac{2}{3}$~ML. We present and discuss our results parallelly to create a picture of the trends for clean Cu(211) and CO molecules adsorbed on this surface. Finally, section \ref{Summary} provides a short summary of the main results and conclusions.

\section{Methodology}
\label{Methodology}

\begin{figure*}[htb]
\psfig{figure=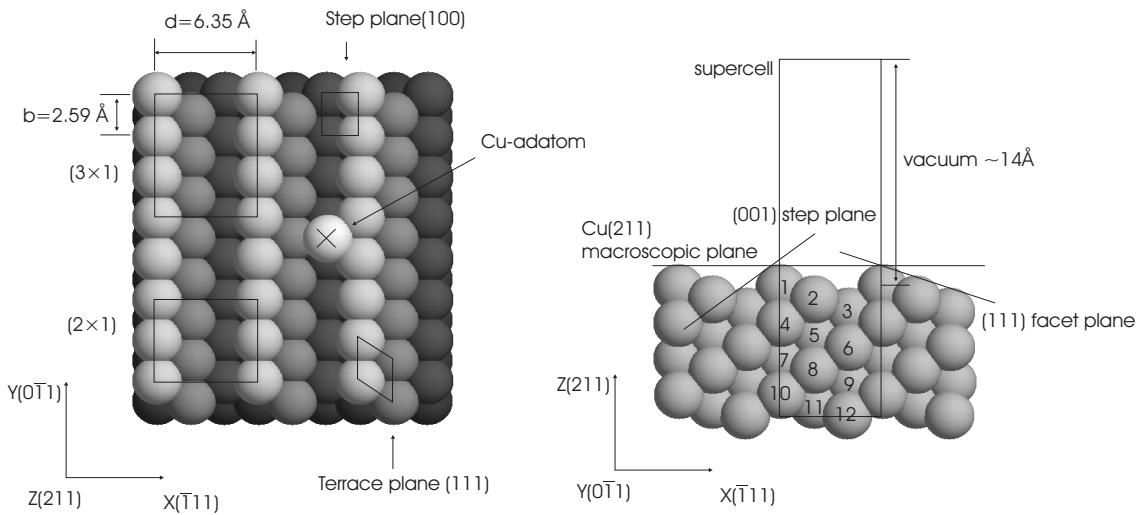,width=15cm,clip=true}
\caption{\label{figure structure 211} Top and side views of vicinal Cu(211)$\equiv$(3(111)$\times$(100)). Typical parallelograms for (111) terraces and rectangles for (100) steps are indicated. 
In addition we show  (2$\times$1) and (3$\times$1) cells for the CO adsorption structures.}
\end{figure*}

In this work, a slab approach is adapted to describe the metallic surface. Twelve (resp. 14 layers for clean surface) (211) crystallographic layers are repeated periodically in a supercell geometry. The vicinal (211) surface consists of (111) oriented terrace planes and monoatomic (100) steps according to the Somorjai notation (3(111)$\times$(100)) \cite{ss:Lang:30}. The macroscopic surface plane is rotated by $19.5^{\circ}$ from the (111) plane. The distance between the steps is $6.25$~\AA~  according to the theoretical bulk lattice parameter a$_{Cu}~=~3.664$~\AA~ of fcc Cu. The structure of the clean Cu(211) surface is shown in Fig. \ref{figure structure 211}. We consider CO molecules in the (2$\times$1) and the (3$\times$1) unit cells. 

The calculations presented in this study are performed within a plane-wave density functional framework. We have used the Vienna Ab-initio Simulation Package (VASP) \cite{vasp,prb:Kresse:54} and employed  projected augmented-wave (PAW) potentials \cite{prb:Blochl:50,prb:Kresse:59}. For exchange and correlation the functional proposed by Perdew and Zunger \cite{prb:Perdew:23} is
used, adding (semi-local) generalized gradient approximation (GGA) of the PW91 \cite{prb:Perdew:46}. Further, the RPBE exchange-correlation functional \cite{prb:Hammer:59} is used, since it is more accurate for the CO adsorption on the Cu(111) surface \cite{jpcm:Gajdos:sub}. It is well known that in some cases DFT calculations fail to produce the correct adsorption site for CO on transition or noble metals surfaces \cite{jpcb:Feibelman:105,jpcm:Gajdos:sub}. It has been shown that this failure arises from the underestimation of the HOMO-LUMO gap of the CO molecule favoring backdonation to the antibonding 2$\pi^{\star}$ state and adsorption in a site with higher coordination \cite{prb:Kresse:68}. Very recently, Kresse et al. \cite{prb:Kresse:68} have shown that the correct site-preference can be achieved by adding a Hubbard-like on-site Coulomb repulsion \cite{prb:Dudarev:57,prb:Liechtenstein:52} to the molecular Hamiltonian of CO leading to an increased HOMO-LUMO gap. We have applied the special GGA+U method \cite{prb:Kresse:68} to examine the effect of stronger electronic correlation on the prediction of the correct adsorption site.

 We have used (4$\times$6$\times$1) and (4$\times$4$\times$1) k-point meshes according to Monkhorst-Pack for the (2$\times$1) and (3$\times$1) unit cells, respectively. The adsorption energy (E$_{ads}$) was defined as the difference between the total energy of the studied system (Cu(211) surface with CO) and the individual components of the system (Cu(211) surface and CO) \cite{ss:Gajdos:531}. 

The atoms in the first five layers and C and O atoms of the CO molecule were allowed to relax. The geometry optimizations were caried according to conjugate-gradient algorithm. 

The finite-difference method is used for the calculation of stretching frequencies. In the finite-difference technique, total energies and Hellmann-Feynman forces are evaluated as a function of the atomic displacements from their equilibrium positions; C and O atoms are moved in the direction of carthesian axis (x,y,z) by 0.04 \AA~and the substrate is frozen. The Hessian dynamical matrix is created from the second derivatives of the total energy with respect to the positions of CO. The matrix is diagonalized and normalized eigenvectors denote modes of the vibrations and corresponding eigenvalues are proportional to stretching frequencies of the vibrational modes \cite{jpcm:KingSmith:2}. 

We have applied a Tersoff-Hamman approach to calculate STM images from the ab-initio calculations where the STM contour plot is approximated by the charge density around the Fermi level ($\pm\sim50$ meV) \cite{prl:Tersoff:50}. Constant current topographs are approximated by constant charge density iso-surfaces. The iso-surfaces are given for the charge density value of e$^{-5}$ eV/$\rm \AA^{3}$.

\section{Clean C$\rm u$(211) surface}
\label{Clean Cu(211) surface}

We have calculated multilayer relaxations of the clean Cu(211) surface and also of clean Cu(211) with one Cu-adatom on the (111) terrace adjacent to a step (see cross in Fig. \ref{figure structure 211} as well as Fig. \ref{figure adatom pictures}, model a). For the calculations, the lateral distance between the extra Cu-adatoms was chosen to be $3$b (see Fig. \ref{figure structure 211}) for the extra Cu atoms in the direction of the steps. The changes of the interlayer spacing from both experiment (LEED \cite{jvsta:Seyller:17}) and theoretical studies are tabulated in Table \ref{table relaxations 211}. Notice that these changes are not strictly oscillatory.
The earlier and current theoretical and experimental studies predict the same trends for the interlayer relaxation (--,--,+) for four surface layers. Such behaviour has been obtained in a FLAPW study \cite{prb:Geng:64}, but the authors report a more pronounced contraction of the first interlayer distance as compared to the experimental results and the present theoretical study. On the other hand, the LDA \cite{prb:Wei:57} and current GGA calculations give very similar results which are moreover close to the experiment. Inclusion of a Cu-adatom next to a step on the Cu(211) surface expands only its local environment which gives rise to a small buckling of the layer below the Cu-adatom.

\begin{table*}[phtb]
\begin{center}
\begin{tabular}{c|lllllll}
\hline
Method & $\Delta d_{12}$ [\%] & $\Delta d_{23}$ [\%] & $\Delta d_{34}$ [\%] & $\Delta d_{45}$ [\%] & $\Delta d_{56}$ [\%] & $\Delta d_{67}$ [\%] & $\Delta d_{78}$ [\%]  \\ 
\hline
this work, Cu(211)                  &--14.4 &--11.4 & +8.1  &--3.0 &--2.8 & 0.0  &       \\ 
this work, Cu(211) with Cu-adatom   &--14.0 &--14.3 &--4.5  & +4.6  &--2.4 &--0.2 &  0.0  \\   
\hline
LDA+norm$^a$ \cite{prb:Heid:65} &--12.2 &--9.5  & +8.7  &--2.1 &--1.6 & +1.5 &--0.1 \\ 
LDA \cite{prb:Wei:57}         &--14.4 &--10.7 & +10.9 &--3.8 &--2.3 & +1.7 &--1.0 \\ 
FPLAPW \cite{prb:Geng:64}     &--28.4 &--3.0  & +15.3 &--6.6 & +0.7 & +3.0 & +0.0  \\ 
EAM$^b$ \cite{prb:Durukanoglu:55} &--10.3 &--5.4  & +7.3  &--5.7 &--1.2 & +4.0 &--2.6 \\ 
EAM \cite{ss:Sklyadneva:416}  &--10.3 &--5.1  & +7.3  & -5.6 &--1.1 &      &        \\ 
\hline
LEED \cite{jvsta:Seyller:17}  &--14.9$\pm$4.1 &--10.8$\pm$4.1 & +8.1$\pm$4.1  &      &      &      &        \\ 
\hline
\end{tabular}
\end{center}
\begin{flushleft} $^a$ Local Density Approximation with with norm-conserving, nonlocal pseudopotentials \\ 
$^b$Embedded-atom method\\ 
\end{flushleft}
\caption{\label{table relaxations 211}
 Relaxations of the 14 layer clean Cu(211) surface and the clean Cu(211) surface with one Cu-adatom on the terrace close to a step (see Fig. \ref{figure adatom pictures}a). The bulk interlayer spacing is d$_{bulk}$~=~$0.748$~\AA~ for Cu(211). Fig. \ref{figure structure 211} indicates the indexing of the individual layers.
}
\end{table*}

The workfunction of clean Cu(211) ($\Phi_{Cu(211)}$) was calculated to be 4.46~eV. It is lower than the workfunction for Cu(111) ($\Phi_{Cu(111)}$=5.00 eV) \cite{jpcm:Gajdos:sub}, in accordance with Smoluchowski's consideration of the smoothening effect of the surface charge density that is responsible for the workfunction reduction on more open surfaces \cite{prb:Smoluchowski:60}. Additional Cu atoms on (111) terraces close to the steps increase the overall dipole moment and consequently decrease the workfunction. Summarizing, the interlayer relaxations are well described by our theoretical tool and the trend (--,--,+) is consistent with experimental observation.

\section{CO adsorption}
\label{CO adsorption}

\subsection{Structural phases}
\label{Structural phases}

We have searched through the literature to determine the most probable adsorption sites for the different coverages. The majority of investigations claim that CO stands more or less upright on the atomic rows forming the steps. As credible sites we therefore considered top and bridge sites on the upper parts of the step edges. We have theoretically examined $\frac{1}{2}$ and $\frac{2}{3}$~ML coverage phases of CO molecules as well as a hypothetical $\Theta=\frac{1}{3}$ ML phase. We found in the literature also chance for CO-adsorption on the (111) surface facet with the CO axis almost parallel to the substrate \cite{jcp:Radnik:110}, but we were not able to observe a minimum of the potential energy surface for this configuration.

\subsection{Structural and energetic properties}
\label{Structural and energetic properties}

\begin{figure*}[phtb]
\psfig{figure=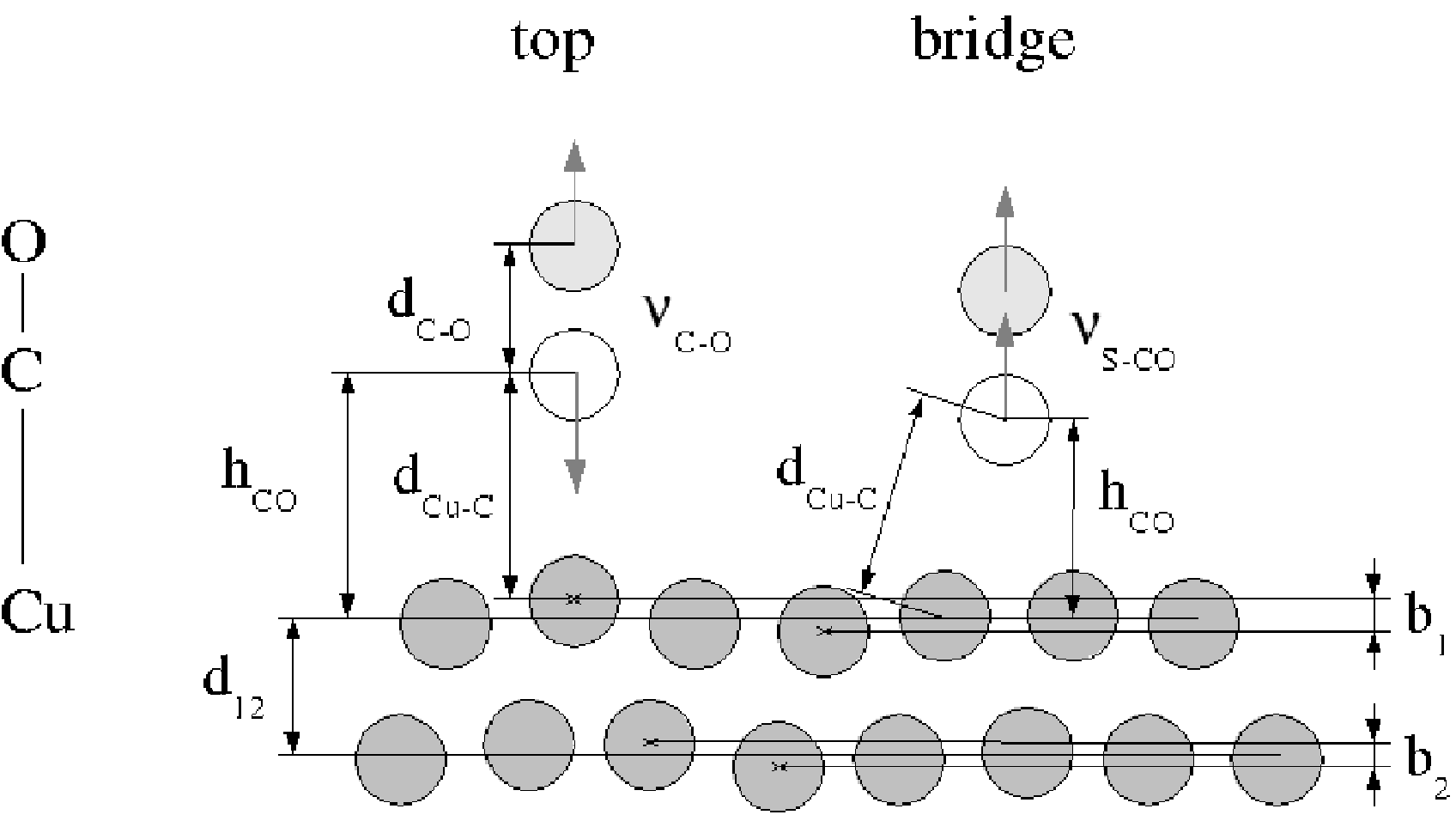,width=14cm,clip=true,angle=0}
\caption{\label{figure sketch properties}
The sketch of the calculated structural and vibrational properties for the top and bridge sites. The reported quantities are: $d_{C-O}$ - carbon-oxygen bond length , $h_{CO}$ - height of CO molecule above the surface (metal surface - carbon distance), $d_{Cu-C}$ - Cu-C bond length, $\Delta d_{12}$, (resp. $\Delta d_{23}$ - average inter-layer spacing between first and second layer), $b_{1}$,$b_{2}$ - buckling of $1^{st}$ and $2^{nd}$ layer, $\nu_{C-O}$ - stretching mode of C-O molecule where arrows denote distorsions, $\nu_{S-CO}$ - S(urface)- CO vibration. 
}
\end{figure*}

Our calculations indicate that CO molecules at the steps adsorb upright with respect to the macroscopic surface plane as long as they do not occupy nearest neighbor sites at the step edges as in one of the possible configurations in the high coverage (3$\times$1) phase. The CO molecules affect only substrate atoms around their local environments and reduce the inward relaxation of atoms. The same trend has been observed for CO adsorption on transition and noble metal surfaces \cite{jpcm:Gajdos:sub}. The sketch of the calculated structural and vibrational properties is shown in Fig. \ref{figure sketch properties}. The values of the structural properties for all calculated adsorption sites are presented in Table \ref{table properties of CO on Cu(211)}. 

\begin{table*}[phtb] 
\begin{center} 
\begin{tabular}{r|rrr|rr|rr} 
 \hline 
structural phase & \multicolumn{3}{c|}{p(3$\times$1)CO}& \multicolumn{2}{c|}{p(2$\times$1)CO} & \multicolumn{2}{c}{p(3$\times$1)2CO} \\
coverage  & \multicolumn{3}{c|}{$\Theta$=$\frac{1}{3}$~ML} & \multicolumn{2}{c|}{$\Theta$=$\frac{1}{2}$~ML} & \multicolumn{2}{c}{$\Theta=\frac{2}{3}$~ML} \\
site(s)  & top(Cu-adatom)  & top & bridge & top & bridge & top-top & top-bridge \\ 
\hline 
$d_{C-O}$ [\AA]      & 1.153 &  1.154&  1.169& 1.154 & 1.168 &  1.156&  1.154,~1.170\\
$h_{CO}$ [\AA]       & 1.84  &  1.85 &  1.51 & 1.84  & 1.50  &  1.81 &  1.84,~1.52\\
$d_{Cu-C}$ [\AA]     & 1.84  &  1.85 &  1.98 & 1.84  & 1.97  &  1.85 &  1.84,~1.97\\
$\Delta d_{12}$ [\%] &--13.9 &--5.7  &--2.8  &--2.9  &  0.9  &  0.9  &  4.8 \\
$b_{1}$ [\AA]        & 0.04  &  0.13 &  0.06 &  0.09 &  0.0  &  0.11 &  0.05 \\
$\Delta d_{23}$ [\%] &--6.3  &--14.4 &--14.3 &--15.9 &--14.9 &--16.4 &--16.0 \\
$b_{2}$ [\AA]        & 0.03  &  0.02 & 0.05  &  0.0  &  0.04 &  0.07 &  0.03 \\
$\Phi$ [eV]          & 4.48  &  4.57 & 4.90  & 4.65  & 5.07  &  4.82 &  5.00 \\ 
\hline 
\end{tabular} 
\end{center} 
 \caption{Calculated structural properties of CO adsorbed at the high symmetry sites (t 
- top, b - bridge) on Cu(211) for coverages varying between $\Theta=\frac{1}{3}$~ML and $\Theta=\frac{2}{3}$~ML. 
The reported quantities are: $d_{C-O}$ - carbon-oxygen bond length , $h_{CO}$ - height of CO molecule above the surface (metal surface - carbon distance), $d_{Cu-C}$ - Cu-C bond length, $\Delta d_{12}$, $\Delta d_{23}$ - change of the average inter-layer spacing, $b_{1}$,$b_{2}$ - buckling of $1^{st}$ and $2^{nd}$ layer, 
$\Phi$ - workfunction. The calculated properties are sketched in Fig. \ref{figure sketch properties}} \label{table properties of CO on Cu(211)} 
\end{table*}

The CO bond length increases with the coordination due to larger occupation of the 2$\pi^{\star}$ orbital \cite{jpcm:Gajdos:sub}. There are only small changes of the CO bond length with increasing coverage.  Furthermore, the CO molecule approaches closer the surface as the coordination increases. The corresponding Cu-C bond length increases with the coordination.  

The bonds that the CO molecules form with Cu atoms reduce the Cu-Cu binding, which is the major reason for the buckling of the first substrate layer. The buckling quickly disappears and is almost zero in the third layer.
The buckling in the layers influences interlayer relaxations in an averaged way; consequently the expansion of the first layer is similar for the bridge site at $\frac{1}{3}$~ML and the top site at $\frac{1}{2}$~ML coverages. The reduction of the second layer is similar for all coverages (except CO adsorption on Cu-adatom). 

The workfunction of clean Cu(211) is $4.46$~eV and increases upon filling top and bridge sites with CO. Moreover, the workfunction increases with CO coverage and also with CO coordination. For the bridge sites the increase in workfunction is by $\sim0.2$~eV larger than for top sites in the (3$\times$1) and by $\sim0.4$~eV for the (2$\times$1) phase. Additionaly, the workfunction for top-top configuration of NO in (3$\times$1) cell is $0.18$ eV lower than the workfunction for top-bridge configuration. This increase may be of further help in determining the optimal adsorption site by respective experiments. 

To determine the most favorable structures from a chosen set of sites at a given coverage we present the adsorption energies in Table \ref{table properties LDAU}. The difference in the adsorption energy between top and bridge site at $\Theta=\frac{1}{3}$~ML and $\frac{1}{2}$~ML is $\sim20$~meV for PW91 exchange-correlation functional, which is small to distinguish the optimal site. 
We observe, for the first time, that the RPBE exchange-correlation functional changes the site preference and gives a different answer concerning the site preference as PW91. The difference in the RPBE  adsorption energy between top and bridge site is not large ($\sim30$~meV), but assigns the top site as the optimal site for $\frac{1}{3}$ and $\frac{1}{2}$ ML coverage. Our conclusions are similar to the earlier semi-empirical calculations of Marinica et al. \cite{ss:Marinica:542}. Earlier work on CO on the close-packed transition and noble metal surfaces led to the conclusion that all exchange-correlation functionals give the same site prediction and the difference among them is mainly in the absolute value of the adsorption energy \cite{jpcm:Gajdos:sub}. 

Underestimation of the one-electron gap between HOMO (5$\sigma$) and LUMO (2$\pi^{\star}$) is the major problem in the theoretical description of CO molecules. The gap can be enlarged by inclusion of an artificial Coulomb repulsion U in the calculation in the way described in the paper of Kresse et al. \cite{prb:Kresse:68,prb:Kohler:sub}. The adsorption energy decreases linearly with the U parameter and the slope increases with coordination. We have taken U~=~$1.0$~eV that leads to the top site as the favorite adsorption site on the Cu(111) surface \cite{Gajdos:prep}, in accordance with experiment. In our calculations this again changes the preference from the bridge site to the top site for the CO adsorption in the hypothetical (3$\times$1) and the (2$\times$1) unit cell. The CO molecule in the on-top site binds stronger by $\sim130$~meV to the substrate. This also indicates that PW91 exchange-correlation functional can not only greatly overestimate the adsorption energy, but can also lead to a qualitatively different result.

\begin{figure*}[phtb]
\psfig{figure=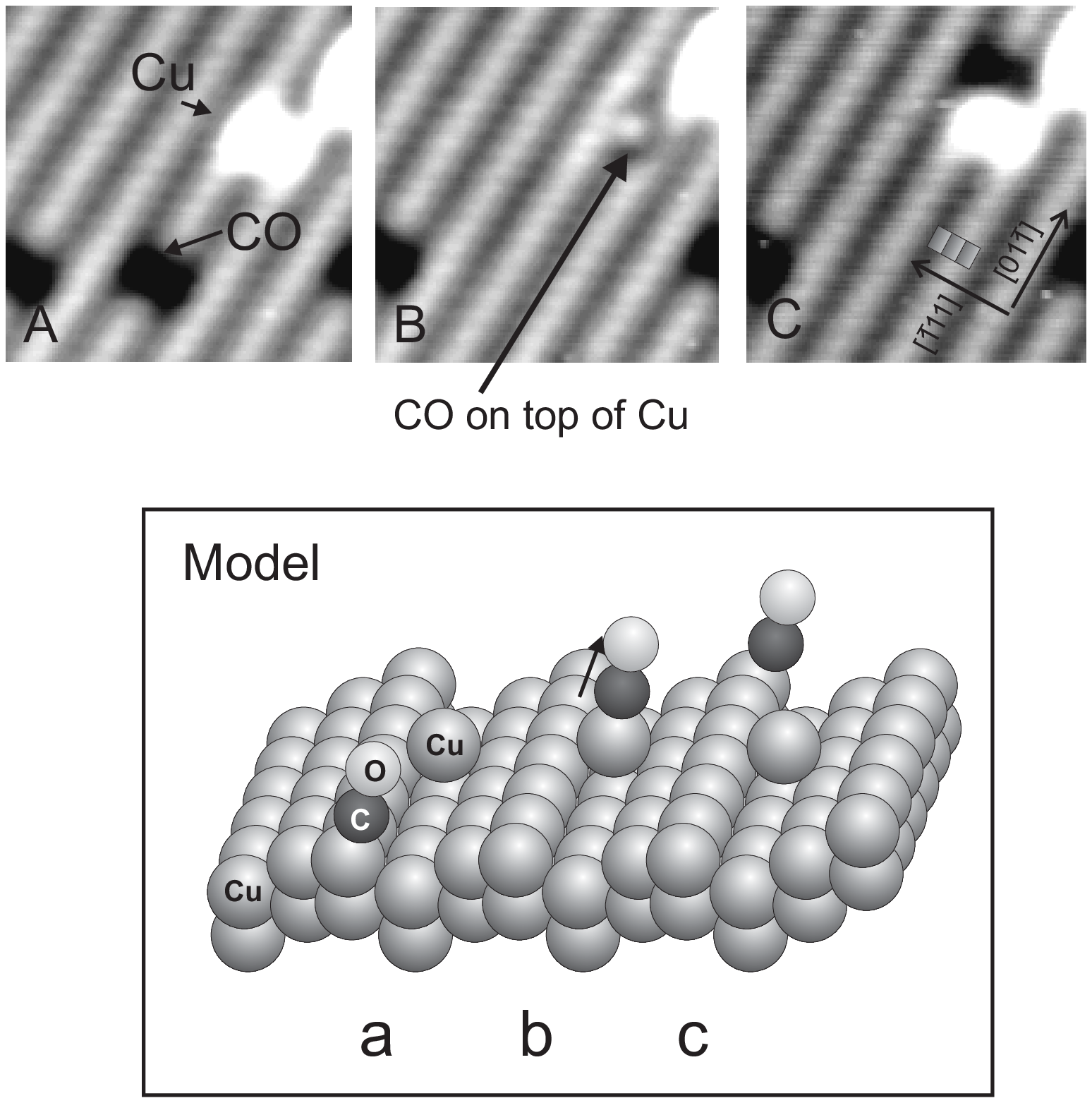,width=14cm,clip=true,angle=0}
\caption{\label{figure adatom pictures}
STM-images and sphere models of atomic and molecular arrangements during an STM experiment, in which a single CO-molecule was pushed along the upper part of a step edge. Notice that a Cu-adatom is located by STM manipulation close to the lower step edge (A,a). Upon manipulation along the upper step edge the CO molecule goes to the Cu-adatom (B,b). A larger force is necessary to move the CO from the Cu-adatom back to the step edge in accordance with present calculations.
}
\end{figure*}

It should be emphasized that the CO molecule on top of an Cu-adatom located at the lower step edge binds stronger by $0.16$~eV for PW91, $0.16$~eV for RPBE and $0.18$~eV for GGA+U than CO at top sites on the steps. This is in accordance with experimental STM manipulation results which are outlined in Fig. \ref{figure adatom pictures}. In Fig. \ref{figure adatom pictures}A a configuration is shown which was created artificially by manipulating a Cu-adatom to an intrinsic step edge of the Cu(211) as depicted in the model \ref{figure adatom pictures}a. A CO molecule located initially on top of an atom at the step was then pushed along the step edge. It was found that the CO goes on top of the Cu-adatom as depicted in Fig. \ref{figure adatom pictures}B and model b. Upon manipulating the CO molecule further to move it back to the step edge (see Fig. \ref{figure adatom pictures}C and model c), it was found that a larger force was required. As direct force measurements with the STM are not possible, the force is measured semi-quantitatively by the manipulation resistivity: Smaller resistivities imply larger forces because the tip is closer to the manipulated particle. The respective tunneling resistivities were $400$ k$\Omega$ and $270$ k$\Omega$. Thus the experimental result indicates an enhanced bonding of the CO to the single Cu-adatom as compared to a CO molecule sitting on top of a Cu-atom in the row of atoms building up the step. This experimental observation is strongly corroborated by the present theoretical results. This result is also of appreciable practical importance as it proves, that the CO molecules transfered deliberately to the tip apex by vertical manipulation \cite{apl:Bartels:71} prefer the adsorption at the most exposed site, i.e. the single metal atom at the tip apex.

\begin{table*}[phtb]
\begin{center} 
\begin{tabular}{r|rrr|rr|rr} 
 \hline 
structural phase & \multicolumn{3}{c|}{p(3$\times$1)CO}& \multicolumn{2}{c|}{p(2$\times$1)CO} & \multicolumn{2}{c}{p(3$\times$1)2CO} \\
coverage & \multicolumn{3}{c|}{$\Theta=\frac{1}{3}$~ML}&  
 \multicolumn{2}{c|}{$\Theta=\frac{1}{2}$~ML} & \multicolumn{2}{c}{$\Theta=\frac{2}{3}$~ML} \\
site(s) & top(Cu-adatom) & top & bridge & top & bridge & top-top & top-bridge \\ 
\hline 
$E_{\rm ads,PW91}$ [eV] & --1.08 & --0.96 & --0.98 & --0.96 & --0.98 & --0.94 & --0.97  \\
$E_{\rm ads,RPBE}$ [eV] & --0.84 & --0.68 & --0.65 & --0.68 & --0.65 & --0.65 & --0.67  \\
$E_{\rm ads,GGA+U}, U=1.0~eV$ [eV] & --0.90 & --0.72 & --0.59 &     --  &  --     & -- & -- \\
 \hline 
\end{tabular} 
\end{center} 
 \caption{Adsorption energies $E_{ads}$ of CO on Cu(211) calculated with PW91, RPBE exchange-correlation functionals and with the GGA+U method.
} \label{table properties LDAU} 
\end{table*}

\subsection{Vibrational properties}
\label{Vibrational properties}

We present the calculated vibrational frequencies of CO molecules adsorbed on Cu(211) in Table \ref{table vibrations CO/Cu(211)}. The C-O stretching frequencies ($\nu_{C-O}$) agree with the already proposed trend that the stretching frequency strongly depends on coordination. In our case the stretching frequency for the on-top site is 2049 cm$^{-1}$ and 1925 cm$^{-1}$ for the bridge site at a coverage of $\frac{1}{3}$~ML, i.e. red-shifted by 87 cm$^{-1}$ and 211 cm$^{-1}$ compared to the calculated gas-phase value of 2136 cm$^{-1}$ \cite{jpcm:Gajdos:sub}.

$\nu_{C-O}$ increases with the coverage for the on-top sites and decreases with increased coordination (bridge sites). A CO molecule adsorbed on-top of the Cu single adatom adjacent to a step edge vibrates in a steep potential as demonstrated by the highest C-O vibrational frequency of 2071 cm$^{-1}$.

\begin{table*}[phtb]
\begin{center} 
\begin{tabular}{r|rrr|rr|rr} 
 \hline 
 structural phase & \multicolumn{3}{c|}{p(3$\times$1)CO}& \multicolumn{2}{c|}{p(2$\times$1)CO} & \multicolumn{2}{c}{p(3$\times$1)2CO} \\
coverage & \multicolumn{3}{c|}{$\Theta=\frac{1}{3}$~ML}&  
 \multicolumn{2}{c|}{$\Theta=\frac{1}{2}$~ML} & \multicolumn{2}{c}{$\Theta=\frac{2}{3}$~ML} \\
site(s) & top(Cu-adatom) & top & bridge & top & bridge & top-top & top-bridge \\ 
\hline 
$\nu_{C-O}$ [cm$^{-1}$]  & 2071  & 2049  & 1925  & 2061  & 1938 & 2061,2007  & 2057,1908\\
$\nu_{S-CO}$ [cm$^{-1}$] & 346   & 333   & 287   & 342   & 291  & 341,339    &  334,303\\
 \hline 
\end{tabular} 
\end{center} 
 \caption{Calculated CO frequencies; symmetric stretching mode ($\nu_{C-O}$ and S(urface)-CO vibration ($\nu_{S-CO}$). There are two vibrational frequencies, in-phase and out-of-phase, for two molecules in the unit cell. 
} \label{table vibrations CO/Cu(211)} 
\end{table*}

Two molecules in the (3$\times$1) unit cell increase the number of possible vibrational modes, adding out-of-phase vibrational frequencies. These frequencies are included as the second value for the top-top and top-bridge configurations in Table \ref{table vibrations CO/Cu(211)}. We observe two important differences between the vibrational frequencies of top-top and top-bridge species in the (3$\times$1) surface unit cell: different splitting of the $\nu_{C-O}$ and $\nu_{S-CO}$ modes. The in-phase symmetric mode $\nu_{C-O}$ for the top-top configuration is calculated to be 2061 cm$^{-1}$. The out-of-phase symmetric mode where both molecules vibrate in opposite directions is 2007 cm$^{-1}$. In the top-bridge configuration the in-phase $\nu_{C-O}$ mode (2057 cm$^{-1}$) is close to the in-phase mode of the top-top configuration, but the out-of-phase symmetric mode is by almost 100 cm$^{-1}$ softer (1908 cm$^{-1}$). 

The experimental EELS study \cite{jcp:Radnik:110} provides values for both in-phase and out-of-phase  CO stretching modes : 2088 cm$^{-1}$ - in-phase and 1881 cm$^{-1}$ - out-of-phase which is already an  indication that the adsorption takes place in the top-bridge configuration. 

The second difference concerns the splitting of the $\nu_{S-CO}$ modes. 
The in-phase and out-of-phase splitting is 2 cm$^{-1}$ for the top-top configuration and 31 cm$^{-1}$ for the top-bridge configuration. The experimental EELS study \cite{jcp:Radnik:110} provides only one S(urface)-CO frequency (341 cm$^{-1}$) for the high-coverage phase with two symmetric modes. In addition to the experimentally observed mode at 341 cm$^{-1}$ we would expect another mode at a lower frequency and with appreciably smaller intensity. This was not observed in Ref. \cite{jcp:Radnik:110} probably due to its small intensity and the large contribution of the specular beam in this frequency region. The $\nu_{S-CO}$ frequency of 306.7 cm$^{-1}$ was calculated for CO in off-top site without phonon bath and the frequency of 343 cm$^{-1}$ with the phonon bath at 100 K \cite{ss:Marinica:497}. The later frequency is in nice agreement with our calculated stretching frequency of 346 cm$^{-1}$ (resp. 342 cm$^{-1}$ for $\frac{1}{2}$ ML coverage) and the experimentally observed frequency (341 cm$^{-1}$).

\subsection{STM pictures}
\label{STM pictures}

Calculated charge density contours have frequently been used for comparison with experimental STM
images and have been able to shed light on complex adsorption systems \cite{prl:Surnev:87}. 

\begin{figure*}[phtb]
\psfig{figure=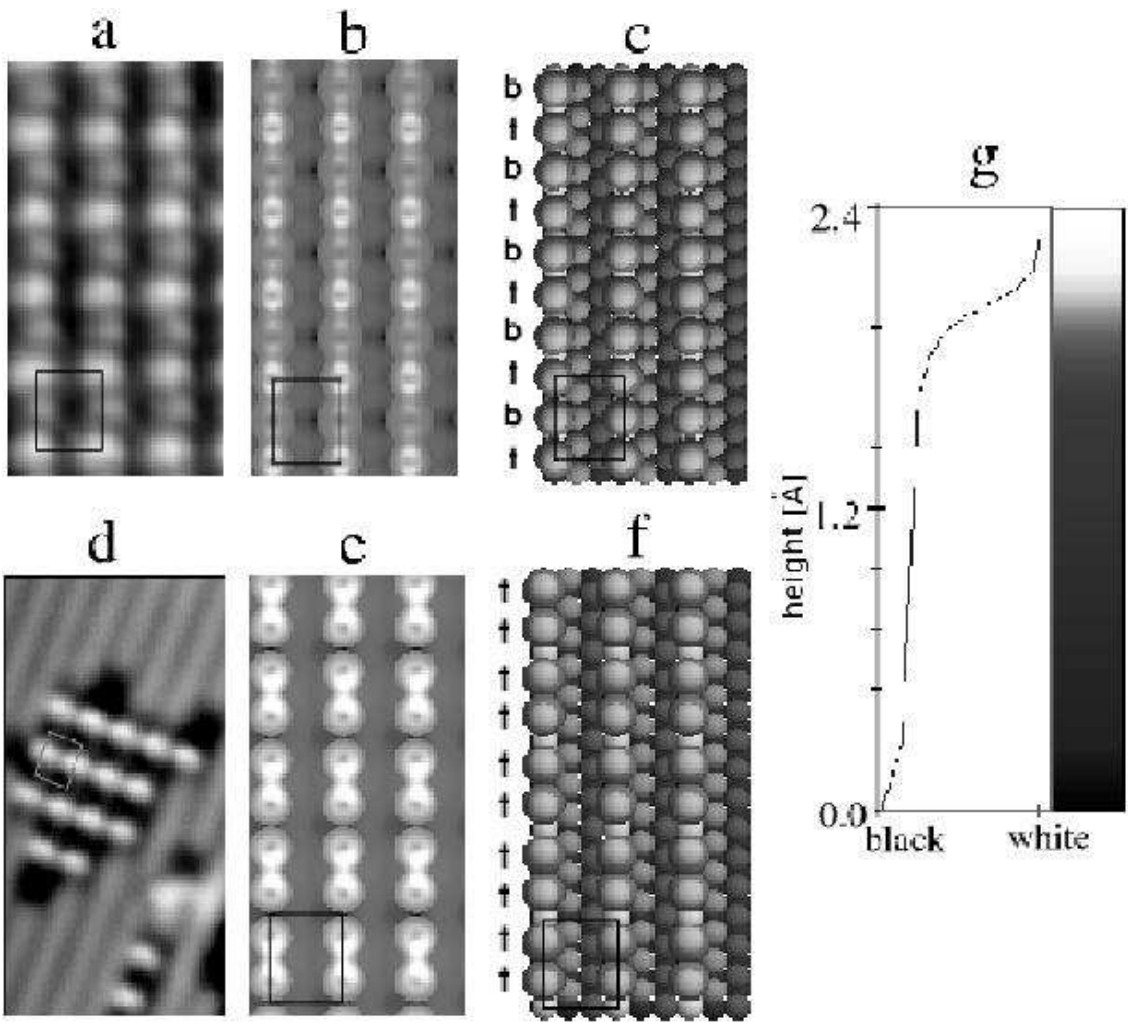,width=15cm,clip=true,angle=0}
\caption{\label{figure STM pictures} Experimental and calculated STM images and corresponding models with two CO atoms in (3$\times$1) cell: (a)  experimental STM image for top-bridge configuration, (b) calculated STM-image for top-bridge, (c) model for top-bridge configuration, (d) experimental STM image for top-top configuration, (e) calculated STM image and (f) model of top-top configuration. The colors of the minimal (0 \AA) and maximal (2.4 \AA) height of the isosurface are given in (g). The minimal height corresponds to the distance of 2.8 \AA~ above the surface for the top-bridge and 2.5 \AA~ for top-top configuration.}
\end{figure*}

In Fig. \ref{figure STM pictures} we compare 
typical experimental STM-pictures (Fig. \ref{figure STM pictures}a and Fig. \ref{figure STM pictures}d) with calculated charge density contours (Fig. \ref{figure STM pictures}b and Fig. \ref{figure STM pictures}e) for both CO-configurations discussed for the (3$\times$1) phase, namely top-bridge and top-top; the corresponding structure models are show in Fig. \ref{figure STM pictures}c and f. The calculated top-bridge arrangement (Fig. \ref{figure STM pictures}b) actually resembles closely the experimental STM-picture (Fig. \ref{figure STM pictures}a). The situation is, however, quite different for the case of the top-top-arrangement (Fig. \ref{figure STM pictures}e and Fig. \ref{figure STM pictures}d): Whereas a double peak structure is seen in the calculations there is only one peak visible in the STM-pictures, whose maximum lies in between the two adjacent CO-molecules.

Furthermore, in most  STM pictures the peak appears elongated in the direction perpendicular to the steps in disagreement with the charge density contour calculations. This is alarming, since this top-top-configuration is unequivocally established and identified in careful STM manipulation and imaging experiments \cite{cpl:Zophel:310}. We can rule out experimental reasons for an explanation of this discrepancy like lateral averaging of the corrugation due to final tip width or thermal effects smearing out the double peak structure due to vibrations of the molecules. The reason for this discrepancy seems more to be caused by physical effects which are connected with delicate problems in imaging CO-molecules on metal surfaces \cite{apl:Bartels:71}: It is well known that with metal tips CO is imaged as depression (see the isolated CO indicated by an arrow in the STM-image of  Fig. \ref{figure adatom pictures}A). If a CO-molecule is deliberately transferred to the tip and CO's at the surface are imaged with a CO-terminated tip, then the molecules appear as protrusions \cite{apl:Bartels:71}. It should be noted that charge density calculations for the hypothetical low coverage (3$\times$1)-phase yield pronounced maxima at the CO-sites. Therefore,  isolated CO-molecules imaged with a CO-terminated tip  resemble more the calculated charge density contour. 

The reason for the fact that with clean metal tips isolated CO's appear as depressions was recently traced back by Nieminen et al. \cite{ss:Nieminen:552} to interference effects between different electron paths passing near and through the CO, which are canceled when the tip also bears a CO-molecule.  The STM-picture of the top-top arrangement in Fig. \ref{figure STM pictures}d was taken with a clean metal tip, but also with a CO-terminated tip the two adjacent molecules appear as a single protrusion. On the other hand, the STM-picture of the extended (3$\times$1)-phase (Fig. \ref{figure STM pictures}a) was highly likely taken with a CO-terminated tip and might therefore look more close to the calculated charge density contour (Fig. \ref{figure STM pictures}e). Nevertheless, in view of the experimental results for the top-top-configuration (Fig. \ref{figure STM pictures}d), we cannot strictly exclude an interpretation of the STM-picture of Fig. \ref{figure STM pictures}a, as consisting of top-top CO-pairs separated by an unoccupied edge site. On the other hand, in view of the present theoretical result, that the top-bridge configuration is slightly energetically favored, it appears also possible that this configuration is thermodynamically favored due to the interplay of energy and configurational entropy effects \cite{cpl:Yoshinobu:211,prl:Grossmann:71}. More experimental data are needed to solve this problem unequivocally. The calculated differences in work function presented here might be of help in this respect. 

\section{Summary}
\label{Summary}
We have used the DFT code VASP to investigate clean and CO-covered Cu(211) vicinal surfaces. We consider the most probable sites according to experiment: on-top, bridge and combinations of these two sites. We show that the clean Cu(211) surface tries to smooth its surface and relaxes inwards in the first and second layers and then it compensates this relaxation by an outward relaxation of the third layer.

CO molecules adsorb perpendicular to the macroscopic surface in on-top sites at low coverages up to $\Theta=\frac{1}{2}$ ML. From energy considerations it is non-trivial to distinguish which phase is the most favorable for $\Theta=\frac{2}{3}$~ML. Our calculated adsorption energies for the CO molecules suggest the top-bridge configuration of the CO molecules on the Cu(211) at this high coverage. However, the top-top configuration is the only one safely established experimentally.

From the vibrational properties the top-bridge configuration appears to be thermodynamically favored, but more experiments are needed to settle this subtle point. Calculations of the CO-adsorption energy on a single Cu-adatom adjacent to a step edge show an increase of $\sim160$ meV (RPBE) with respect to CO-adsorption on Cu-atoms embedded in the step edges. This is in agreement with experimental observations using lateral manipulation of the CO. This result is very important for vertical manipulation as it shows that CO prefers to adsorb at low coordination sites as available at the tip apex.



\end{document}